\input{aipcheck}

\documentclass[
    ,final            
  ]
  {aipproc}

\layoutstyle{6x9}

\newcommand{\chandra}{{\it Chandra} }
\newcommand{\xmm}{{\it XMM-Newton} }

\begin{document}

\title{Fossil Galaxy Groups -- Ideal Laboratories for Studying the Effects of AGN Heating.}


\author{Nazirah N. Jetha}{
  address={Department of Physics, University of Alabama, Huntsville, AL, 35899}
}

\author{Habib Khosroshahi}{
  address={School of Astronomy, Institute for Research in Fundamental Sciences
(IPM), P. O. Box 19395-5531
Tehran, Iran}
}

\author{Somak Raychaudhury}{
  address={School of Physics and Astronomy, University of Birmingham, UK}
}

\author{Chandreyee Sengupta}{
  address={NCRA-TIFR, Pune, India}
}

\author{Martin Hardcastle}{
  address={University of Hertfordshire, UK}
}

\begin{abstract}
 We present the first of a sample of fossil galaxy groups with
 pre-existing {\emph{Chandra}} and/or \emph{XMM-Newton} X-ray
 observations and new or forthcoming low frequency GMRT data --
 RXJ1416.4+2315 (z=0.137). Fossil galaxy groups are ideal laboratories
 for studying feedback mechanisms and how energy injection affects the
 IGM, since due to the lack of recent merging activity, we expect the
 IGM to be relatively pristine and affected only by any AGN activity
 that has occurred in the group. Our Chandra X-ray observations reveal
 features resembling AGN-inflated bubbles, whilst our GMRT radio data
 show evidence of extended emission from the central AGN that may be
 filling the bubble. This has enabled us to estimate the work done by
 the central AGN, place limits on the rates of energy injection and
 discuss the nature of the plasma filling the bubble.
\end{abstract}

\maketitle


\section{Introduction}
Outbursts from active galactic nuclei (AGN)are the
most likely method of energy injection into galaxy groups and clusters
to prevent catastrophic cooling of the intergalactic medium (IGM)
(\cite{2002MNRAS.332..271R}, \cite{2006MNRAS.373..739N}, \cite{2002ApJ...573..542B} for simulations; \cite{2004ApJ...607..800B},
\cite{2005ApJ...625..748C} for observations). Their
cyclic nature allows them to be triggered when the IGM is cooler, and for the outflow to terminate when the gas
has been sufficiently heated
(e.g. \cite{Jetha:2008p629}, \cite{2000ApJ...543..611O}, \cite{2003MNRAS.344L..43F}, \cite{2007ApJ...659.1153W}).
There is evidence to suggest that the AGN and IGM interact in a
complex feedback loop \cite[e.g.][]{Jetha:2008p629}, whereby whether
the AGN is active or not is controlled by the thermal state of the
IGM, and the energy output is regulated by the amount of cool gas that
accretes onto the AGN.

In many cases, the AGN inflates bubbles into the IGM, seen as
depressions in the X-ray surface brightness.  The thermal properties of gas in and surrounding these bubbles allow us to estimate to within an order of magnitude,
just how much work the AGN does on its surroundings
(see \cite{2005MNRAS.364.1343D}, \cite{2004ApJ...607..800B}
and \cite{Birzan:2008p1127} amongst others).  However, in most groups and clusters on-going galaxy-galaxy mergers may also trigger
AGN.  For this reason, it is not clear that every instance of relic
AGN activity in a group/cluster is feedback triggered,
making it difficult to accurately investigate how the environment
triggers, and is affected by the AGN.

Fortunately, fossil galaxy groups provide an ideal laboratory in which
to study the effects of AGN heating.  These groups are dominated by a
single luminous elliptical galaxy at the center of an extended X-ray
halo similar to those seen in bright X-ray groups. The X-ray emission
in fossils is regular and symmetric, indicating the absence of recent
galaxy merging.  Observationally a galaxy group is classified as a
fossil if it has an X-ray luminosity of $L_{X,bol} \geq 10^{42}
h^{-2}_{50}$ erg s$^{-1}$ that is spatially extended to $\sim$ 100s of
kiloparsecs, and the dominant galaxy is at least 2 R-band magnitudes
brighter than the second ranked galaxy within half the projected
virial radius of the group \cite{jones03}.  The dominant galaxy tends
to be a giant elliptical galaxy with an optical luminosity similar to
BCGs. The observed properties of fossils suggest that an overwhelming
majority of these must be early-formed systems, and have a higher dark
matter concentration compared to non-fossil groups and clusters of
comparable mass \cite{kmpj06}\cite{kpj06},\cite{2007MNRAS.382..433D}.

The early formation epoch implies sufficient time for a cool core to
have developed, and provides an ideal environment to study how
feedback mechanisms work, since the lack of recent merger activity
implies that any visible AGN activity, even if relic activity, should
be purely feedback-driven.  This makes fossil groups ideal
laboratories to study galaxy evolution and IGM heating in the absence
of recent mergers.  Here, we present the results of a preliminary
multi-wavelength study utilizing Giant Metrewave Radio Telescope
(GMRT), \chandra\ and \xmm\ observations, of the fossil galaxy group
RX~J1416.4+2315, which is at $z=0.137$ ($\mathrm{kpc/arcsec} = 2.4$).
We discuss the implications for the nature of the radio plasma, and
the energetics required for the heating model.

\section{The X-Ray and Radio Data}

\subsection{X-Ray Data}
An X-ray analysis of the group was performed and reported in 
\cite{kmpj06}; it was observed both by the \chandra\ and \xmm\
X-ray telescopes, for 15 and 9~ks respectively, and has an X-ray
luminosity within the cooling radius of $L_{\mathrm{X}}=2\times
10^{42}~\mathrm{erg\ s^{-1}}$.  We used the \xmm\ data to obtain
density, temperature and pressure profiles for the group out to large
radii (as shown in \cite{kpj06}), and use these profiles to estimate
the work done by the AGN on the IGM.
  
\subsection{Radio Data}
\begin{figure}
  \includegraphics[height=.25\textheight]{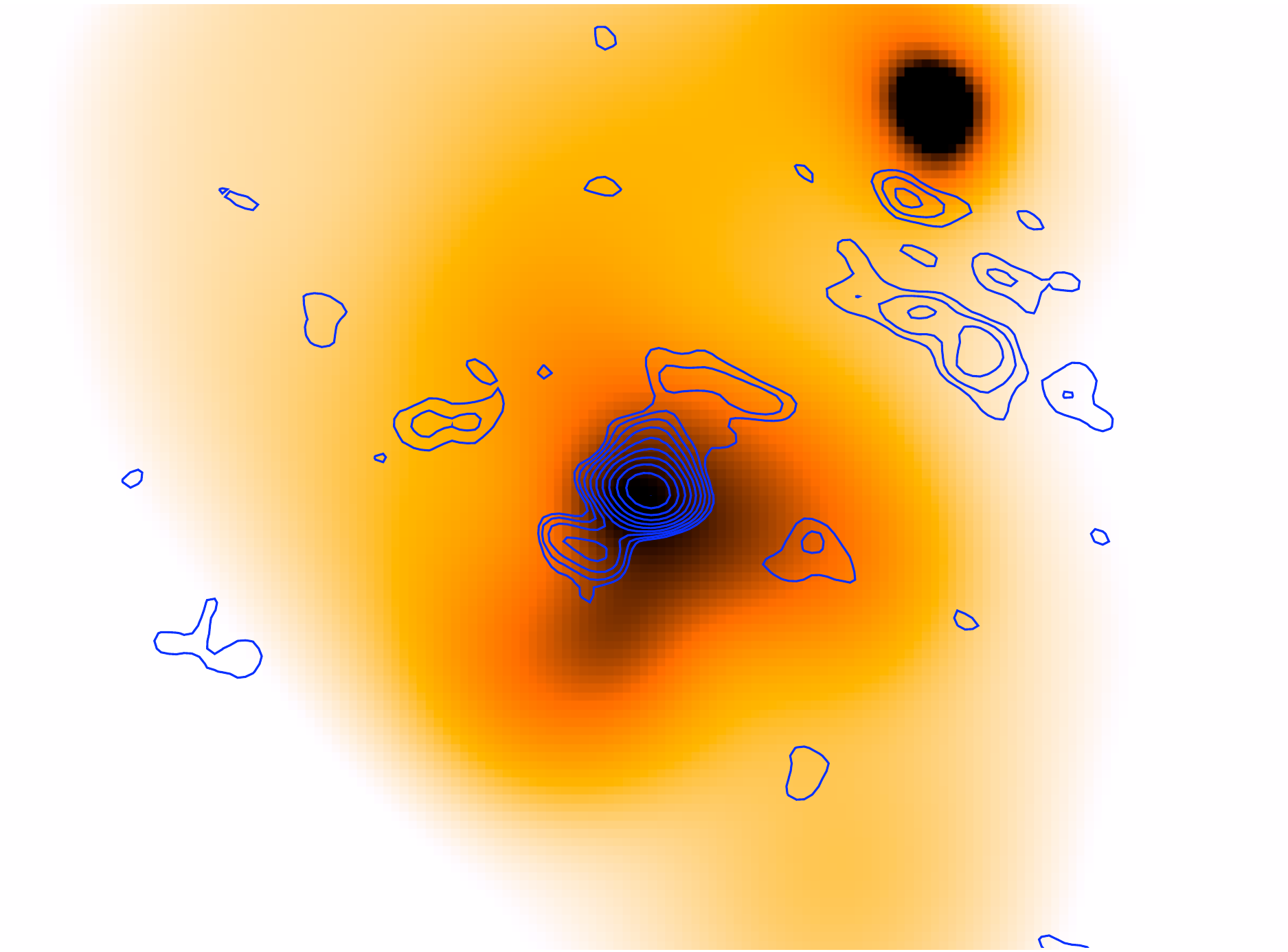}
  \caption{\xmm\ false colour image overlaid with 610~MHz GMRT contours.  The radio extension to the SE co-incides with a depression in the X-ray surface brightness, which we assume is induced by the AGN outflow.}
\label{overlay}
\end{figure}

The group was also observed with the GMRT at 610 and 1420~MHz, and
the data reduced as described in Khosroshahi {\it el al.\ }(in prep.\
).  At 1420~MHz we detect a central point source with no extended
emission, whilst at 610~MHz we detect the central point source
together with an extension to the SE that corresponds to a depression
in the X-ray surface brightness (see Fig.~\ref{overlay}).  Given the
feature's extended nature, spatial coincidence with the surface
brightness feature, and steep spectral index, we assume this to be a
relic bubble feature that is created and possibly filled by the SE radio jet.

\section{Energetics of the Source}
Assuming that the bubble was inflated {\it in situ}, at a sub-sonic
rate, we calculate that the bubble must be approximately $7\times
10^7$~years old.  Then, if the bubble has done $PdV$ work against the
the IGM, we calculate, using the pressure profile calculated
in \cite{kpj06}, that the radio source has done $8.8\times
10^{50}~\mathrm{erg}$ of work over the lifetime of the source,
corresponding to an average energy injection rate of $\sim 4\times
10^{43}~\mathrm{erg\ s^{-1}}$.  This is comparable to, but an order of magnitude higher than the X-ray luminosity of
the source within the radius affected by the bubble, $2\times
10^{42}~\mathrm{erg\ s^{-1}}$, and suggests an inefficient coupling of
energy.  Further, using the central density of the X-ray emitting
medium, we calculate that if the AGN was powered by Bondi accretion,
then the Bondi power, $P_{\mathrm{BONDI}} \sim 8\times
10^{42}~\mathrm{erg\ s^{-1}}$, suggesting that a feedback-driven
accretion mechanism is indeed plausible \cite[see also][]{Jetha:2008p629}.
\section{Physical Nature of the Plasma}
If the bubble is indeed the result of a fairly young AGN outflow, as
suggested by the dynamical age, then  there are two possible
explanations for the physical conditions in the plasma filling the
bubble, and for the steep-spectrum nature of the source.  Either, the
bubble plasma is partially composed of very hot gas, which may leave a
signature in the X-ray, or the plasma is non-thermal in origin and
simply deviates strongly from equipartition.  To test the first case,
we follow the method described in \cite{2008MNRAS.384.1344J}, and find that the X-ray
spectrum of the bubble requires no extra thermal component above the
ambient IGM unless $k_{\mathrm{B}}T_{\mathrm{plasma}}>>kk_{\mathrm{B}}T_{\mathrm{IGM}}$,
suggesting this scenario is unlikely.

In the second case, either that the magnetic field is
significantly different from the equipartition field, or the
plasma is dominated by non-radiating particles.  Using inverse
Compton limits from the X-ray spectrum, together with the radio
measurements presented here, we find that either the plasma must be
magnetically dominant with the magnetic field being two orders of
magnitude greater than the equipartition field; or the plasma
dominated by non-radiating particles, with the ratio of radiating to
non-radiating particles, $k\geq 100$, \cite[see also][]{Birzan:2008p1127}.

\section{Conclusions}

We have presented an X-ray and radio study of the fossil galaxy
group RXJ~1416.4+2315, and have shown the potential of using a sample of
fossil galaxy groups to investigate feedback-driven AGN
activity.  For the case of RXJ~1416.4+2315, simple Bondi accretion,
suggesting a feedback-driven scenario, can account for the current
outburst and the work done on the IGM.  Using limits from the X-ray
and radio data we have shown that the plasma is likely dominated by
non-radiating particles, indicating that it is likely that the jets in
these sources are hadronic in nature (see also \cite{2008MNRAS.386.1709C}).  Further work will concentrate on using
new deep GMRT observations, combined with the existing X-ray data, to
further investigate heating and the nature of the plasma in a larger
sample of fossil groups.


\begin{theacknowledgments}
NNJ would like to thank NCRA-TIFR for their hospitality during her visit there, and Massimiliano Bonamente and Marshall Joy for useful discussions regarding this work.  
\end{theacknowledgments}



\bibliographystyle{aipproc}   

\bibliography{jetha_n}

\begin{thebibliography}{17}
\expandafter\ifx\csname natexlab\endcsname\relax\def\natexlab#1{#1}\fi
\providecommand{\enquote}[1]{``#1''}
\expandafter\ifx\csname url\endcsname\relax
  \def\url#1{\texttt{#1}}\fi
\expandafter\ifx\csname urlprefix\endcsname\relax\def\urlprefix{URL }\fi
\providecommand{\eprint}[2][]{\url{#2}}

\bibitem[Reynolds et~al.(2002)]{2002MNRAS.332..271R}
C.~S. Reynolds, S.~Heinz, and M.~C. Begelman, \emph{MNRAS} \textbf{332},
  271--282 (2002).

\bibitem[Nusser et~al.(2006)]{2006MNRAS.373..739N}
A.~Nusser, J.~Silk, and A.~Babul, \emph{MNRAS} \textbf{373}, 739--746 (2006).

\bibitem[Brighenti and Mathews(2002)]{2002ApJ...573..542B}
F.~Brighenti, and W.~G. Mathews, \emph{The Astrophysical Journal} \textbf{573},
  542 (2002).

\bibitem[B{\^\i}rzan et~al.(2004)]{2004ApJ...607..800B}
L.~B{\^\i}rzan, D.~A. Rafferty, B.~R. McNamara, M.~W. Wise, and P.~E.~J.
  Nulsen, \emph{ApJ} \textbf{607}, 800--809 (2004).

\bibitem[Clarke et~al.(2005)]{2005ApJ...625..748C}
T.~E. Clarke, C.~L. Sarazin, E.~L. Blanton, D.~M. Neumann, and N.~E. Kassim,
  \emph{ApJ} \textbf{625}, 748--753 (2005).

\bibitem[Jetha et~al.(2008{\natexlab{a}})]{Jetha:2008p629}
N.~Jetha, M.~Hardcastle, T.~Ponman, and I.~Sakelliou, \emph{Monthly Notices of
  the Royal Astronomical Society} \textbf{391}, 1052--1062
  (2008{\natexlab{a}}).

\bibitem[Owen et~al.(2000)]{2000ApJ...543..611O}
F.~N. Owen, J.~A. Eilek, and N.~E. Kassim, \emph{ApJ} \textbf{543}, 611--619
  (2000).

\bibitem[Fabian et~al.(2003)]{2003MNRAS.344L..43F}
A.~C. Fabian, J.~S. Sanders, S.~W. Allen, C.~S. Crawford, K.~Iwasawa, R.~M.
  Johnstone, R.~W. Schmidt, and G.~B. Taylor, \emph{MNRAS} \textbf{344},
  L43--L47 (2003).

\bibitem[Wise et~al.(2007)]{2007ApJ...659.1153W}
M.~W. Wise, B.~R. McNamara, P.~E.~J. Nulsen, J.~C. Houck, and L.~P. David,
  \emph{ApJ} \textbf{659}, 1153--1158 (2007).

\bibitem[Dunn et~al.(2005)]{2005MNRAS.364.1343D}
R.~J.~H. Dunn, A.~C. Fabian, and G.~B. Taylor, \emph{MNRAS} \textbf{364},
  1343--1353 (2005).

\bibitem[B{\^\i}rzan et~al.(2008)]{Birzan:2008p1127}
L.~B{\^\i}rzan, B.~R. McNamara, P.~E.~J. Nulsen, C.~L. Carilli, and M.~W. Wise,
  \emph{ApJ} \textbf{686}, 859 (2008).

\bibitem[Jones et~al.(2003)]{jones03}
L.~R. Jones, T.~J. Ponman, A.~Horton, A.~Babul, H.~Ebling, and D.~J. Burke,
  \emph{MNRAS} \textbf{343}, 627 (2003).

\bibitem[Khosroshahi et~al.(2006{\natexlab{a}})]{kmpj06}
H.~G. Khosroshahi, B.~J. Maughan, T.~J. Ponman, and L.~R. Jones, \emph{MNRAS}
  \textbf{369}, 1211 (2006{\natexlab{a}}).

\bibitem[Khosroshahi et~al.(2006{\natexlab{b}})]{kpj06}
H.~Khosroshahi, T.~J. Ponman, and L.~R. Jones, \emph{Monthly Notices of the
  Royal Astronomical Society} \textbf{377}, 595 (2006{\natexlab{b}}).

\bibitem[Dariush et~al.(2007)]{2007MNRAS.382..433D}
A.~Dariush, H.~G. Khosroshahi, T.~J. Ponman, F.~Pearce, S.~Raychaudhury, and
  W.~Hartley, \emph{MNRAS} \textbf{382}, 433 (2007).

\bibitem[Jetha et~al.(2008{\natexlab{b}})]{2008MNRAS.384.1344J}
N.~N. Jetha, M.~J. Hardcastle, A.~Babul, E.~O'Sullivan, T.~J. Ponman,
  S.~Raychaudhury, and J.~Vrtilek, \emph{MNRAS} \textbf{384}, 1344--1354
  (2008{\natexlab{b}}).

\bibitem[Croston et~al.(2008)]{2008MNRAS.386.1709C}
J.~H. Croston, M.~J. Hardcastle, M.~Birkinshaw, D.~M. Worrall, and R.~A. Laing,
  \emph{MNRAS} \textbf{386}, 1709 (2008).

\end{thebibliography}

\IfFileExists{\jobname.bbl}{}
 {\typeout{}
  \typeout{******************************************}
  \typeout{** Please run "bibtex \jobname" to optain}
  \typeout{** the bibliography and then re-run LaTeX}
  \typeout{** twice to fix the references!}
  \typeout{******************************************}
  \typeout{}
 }

\end{document}